\documentstyle[a4,amssymb,prb,aps,epsf]{revtex}

\renewcommand{\a}{\ddot{x}}
\renewcommand{\v}{\dot{x}}

\newcommand{\tf}{\tilde{f}}
\newcommand{\tv}{\tilde{\v}}
\newcommand{\tx}{\tilde{x}}

\newcommand{\lhf}{$\lambda = \frac{1}{2}$}
\newcommand{\lon}{$\lambda = 1$}

\newcommand{\eij}{{\bf\hat{e}}_{ij}}

\newcommand{\ldashed}{--$\,$--}
\newcommand{\sdashed}{-$\,$-$\,$-}
\newcommand{\dotdash}{--$\,\cdot{}\,$--}

\begin{document}

\title{Finite-difference methods for simulation models incorporating
  non-conservative forces}

\author{Keir E. Novik} 
\address{Theory of Condensed Matter, Cavendish Laboratory, University
  of Cambridge, Madingley Road, Cambridge, CB3 0HE, U.K. \\
  {\tt ken21@cam.ac.uk}}

\author{Peter V. Coveney}
\address{Schlumberger Cambridge Research, High Cross, Madingley Road,
  Cambridge, CB3 0EL, U.K. \\
  {\tt coveney@cambridge.scr.slb.com}}

\date{July 29, 1998}

\maketitle

\begin{abstract}
  We discuss algorithms applicable to the numerical solution of
  second-order ordinary differential equations by finite-differences.
  We make particular reference to the solution of the dissipative
  particle dynamics fluid model, and present extensive results
  comparing one of the algorithms discussed with the standard method
  of solution.  These results show the successful modeling of phase
  separation and surface tension in a binary immiscible fluid mixture.
\end{abstract}

\section{Introduction}

Computer simulations are a test-bed for theory, permitting experiments
over which we have nearly complete control.  The effectiveness of
these computer experiments is limited by the power of available
computers, and by their inherently discrete nature since the physical
and chemical phenomena we study are continuous.  Fortunately,
mathematics has provided us with a framework into which we can place
this problem: finite-differences and their continuous analogy,
differential equations.

In the present paper, we shall be concerned with algorithms for
solving second-order ordinary differential equations of the form
\begin{equation}\label{ode}
  \a = f\left(x,~\v\right),
\end{equation}
where $x$ denotes a scalar or vector position, and dots indicate
derivatives with respect to time.  In particular, we are concerned
with determining the motion of a system of particles moving according
to the dissipative particle dynamics (DPD) computational fluid
model\cite{hoogerbrugge:dpd} (see Section~\ref{dpd}).  The large
number of equations and their non-linear nature make direct analytic
mathematical solution infeasible, leaving numerical solution by
finite-differences as the most appropriate approach.  The forces in
our computational fluid model are unusual in that they have a
stochastic component, which depends on the size of the timestep in our
finite-difference algorithm.

Traditional methods for solving Eq.~(\ref{ode}) can be found in
textbooks on numerical analysis,
\cite{burden:analysis,gear:odes,press:nr} and in papers and texts on
molecular dynamics and related computational methods.
\cite{allen:chemical,allen:liquids,berendsen:algorithms,groot:bridging,haile:md,hockney:particles}
Algorithms from numerical analysis textbooks are usually unsuitable
for our purpose because of the unstable nature of the motion of a
many-body system, or an assumption that the forces do not depend on
the timestep.  Most of the other algorithms we find assume
conservative forces, i.e.\ $\a = f\left(x\right)$.

Throughout this paper, we use the notation $x_n = x(t)$ and $x_{n+1} =
x(t+h)$ for the value of a variable (vector or scalar) at successive
timesteps, where $h>0$ is the timestep width.  Variables with tildes
denote temporary quantities, vectors are in bold, and matrices are in
a sans serif font.

\section{Dissipative particle dynamics}\label{dpd}

Hoogerbrugge and Koelman proposed DPD\cite{hoogerbrugge:dpd} as a
novel particulate model for the simulation of complex fluid behavior.
This model is particularly well-suited to model multiphase flows, flow
in porous media, colloidal
suspensions,\cite{boek:colloidal,boek:coll2} microemulsions, and
polymeric fluids.\cite{schlijper:polymer} The traditional approach of
continuum fluid mechanics has met with limited success, and so many
new micro- and mesoscopic approaches have been considered.  In
principle, molecular dynamics (MD) is the most accurate microscopic
approach, although in practice it is too slow in both its quantum
(Car-Parinello) and classical forms because of its excessive detail.
Discrete methods developed from lattice-gas automata (LGA) have had
some success, but they too have problems, such as lacking Galilean
invariance.\cite{bib:bt,bib:dhl}

DPD was developed in an attempt to capture the best aspects of MD and
LGA\@.  It avoids the lattice-based problems of LGA, yet maintains an
elegant simplicity and larger scale that keeps the model much faster
than MD\@.  This simplicity also makes DPD highly extensible, such as
for including the interactions of complex molecules or modeling flow
in an arbitrary number of spatial dimensions.  The key features of the
basic model are that the fluid is grouped into packets, termed
``particles'', and that mass and momentum are conserved but energy is
not.  Particle positions and momenta are real variables, and are not
restricted to a grid.

There are two sequential steps to the action of the original model
proposed by Hoogerbrugge and Koelman:\cite{hoogerbrugge:dpd} (i) an
infinitesimally-short impulse step
\begin{equation}
  {\bf p}_{i,n+1} = {\bf p}_{i,n} + \sum_{j \ne i}\Omega_{ij} \eij,
\end{equation}
and (ii) a propagation step taking time $h$
\begin{equation}
  {\bf x}_{i,n+1} = {\bf x}_{i,n} + {h \over m} {\bf p}_{i,n+1},
\end{equation}
where ${\bf p}_i$ and ${\bf x}_i$ are the momentum and position of
particle $i$, $\eij$ is the unit vector pointing from particle $j$ to
particle $i$, $m$ is the mass of each particle, and
\begin{equation}
  \Omega_{ij} = {3m \left( 1 - {\displaystyle r_{ij} \over\displaystyle
        r_c} \right) \over \pi r_c^2 \rho} \left[ \Pi_{ij} - \omega
    \left({\bf p}_i - {\bf p}_j \right) \cdot{}\eij \right]
\end{equation}
within the cut-off radius $r_c$.  The quantity $r_{ij}$ denotes the
distance separating particles $i$ and $j$, and $\rho$ is the mass
density; note also that the normalization given is correct only in two
dimensions.  The random variable $\Pi_{ij}$ represents the
conservative and stochastic effect of the collisions and gives rise to
fluid pressure, while the second, dissipative, term inside the square
brackets yields fluid viscosity.

Espa\~nol and Warren's analysis\cite{espanol:dpd} showed that the
original DPD model does not satisfy detailed balance, so the
equilibrium states (if they exist) cannot be simply characterized.
Detailed balance is the condition equating the rates of forward and
backward transition probabilities in a dynamical system, and is a
sufficient (but not necessary) condition guaranteeing that the system
has a (Gibbsian) equilibrium
state.\cite{gardiner:stochastic,risken:fokker} Espa\~nol and Warren
formulated a Fokker-Planck equation and equivalent set of stochastic
differential equations which lead to a similar model,
\begin{equation}\label{SDE}
  \cases{%
    d{\bf p}_i = \displaystyle\sum_{j \ne i} {\bf F}_{ij} dt =
    \displaystyle\sum_{j \ne i} \left[ {\bf F}^C_{ij} dt + 
      {\bf F}^D_{ij} dt + {\bf F}^R_{ij} dW_{ij} \right] \smallskip\cr
    d{\bf x}_i = {\displaystyle {\bf p}_i \over\displaystyle m_i}
    dt. 
    } 
\end{equation}
In these equations, ${\bf p}_i$, ${\bf x}_i$, and $m_i$ denote the
momentum, position, and mass of particle $i$, and ${\bf F}^C_{ij}$ is
a conservative force acting between particles $i$ and $j$ while ${\bf
  F}^D_{ij}$ and ${\bf F}^R_{ij}$ are the dissipative and random
forces; $dW_{ij} = dW_{ji}$ are independent increments of a Wiener
process.  By It\^o calculus
\begin{equation}
  dW_{ij} dW_{kl} = \left( \delta_{ik}\delta_{jl} +
    \delta_{il}\delta_{jk} \right) dt ,
\end{equation}
and so $dW_{ij}$ is an infinitesimal of order $\frac{1}{2}$ and we can
write $dW_{ij}=\theta_{ij}\sqrt{dt}$, where $\theta_{ij}=\theta_{ji}$
is a random variable with zero mean and unit
variance.\cite{gardiner:stochastic} With an appropriate choice for the
form of the forces we find detailed balance is satisfied by this
continuous-time version of DPD,\cite{bib:espan95} and so equilibrium
states are guaranteed to exist and be Gibbsian.  To ensure that the
associated fluctuation-dissipation theorem holds, the forces assume the
following forms:\cite{espanol:dpd}
\begin{equation}\label{Fc}
  {\bf F}^C_{ij} = \alpha \omega_{ij} \eij,
\end{equation}
\begin{equation}\label{Fd}
  {\bf F}^D_{ij} = - \gamma \omega^2_{ij} \left( \eij \cdot{}
    {\bf v}_{ij} \right) \eij ,  
\end{equation}
and
\begin{equation}\label{Fr}
  {\bf F}^R_{ij} = \sigma \omega_{ij} \eij,
\end{equation}
where ${\bf v}_{ij} = {\bf p}_i / m_i - {\bf p}_j / m_j$ is the
difference in velocities of particles $j$ and $i$, $\eij$ is the unit
vector pointing from particle $j$ to particle $i$, and $\omega_{ij}$
is a weighting function depending only on the distance separating
particles $i$ and $j$.  The constants $\alpha$, $\gamma$, and $\sigma$
are chosen to reflect the relative importance of the conservative,
dissipative (viscous), and random components in the fluid of interest.
As a consequence of detailed balance and the fluctuation-dissipation
theorem, $\gamma$ and $\sigma$ are related to Boltzmann's constant
$k_B$ and the equilibrium temperature $T$ by
\begin{equation}
  {\sigma^2 \over \gamma} = 2 k_B T.
\end{equation}
In order to remain as close as possible to the original DPD model, we
choose the friction weight function to be
\begin{equation}
  \omega_{ij} = 1 - {\displaystyle r_{ij} \over\displaystyle r_c}
\end{equation}
within the constant cutoff length $r_c>0$, where $r_{ij}$ is the
distance between particles $i$ and $j$.  Adding
Eqs.~(\ref{Fc})--(\ref{Fr}), the total force is
\begin{equation}\label{dpd_force}
  {\bf F}_{ij} = \left[ \alpha - \gamma \omega_{ij} \left( \eij
      \cdot{}{\bf v}_{ij} \right) + {\sigma \theta_{ij} \over
      \sqrt{dt}} \right] \omega_{ij} \eij .
\end{equation}

In summary, the main changes from the original DPD model are the
specification of the motion in terms of differential equations instead
of an impulse step followed by coasting; the separation of the
conservative, dissipative, and random forces, with the strength of
each being controlled by the new constants $\alpha$, $\gamma$, and
$\sigma$; the insertion of an extra factor of $\omega_{ij}$ in front
of the dissipative force; the specification of the thermodynamic
temperature in terms of the model parameters; and the square root
dependence of the random force on size of timestep.

In order to model binary immiscible fluids, we adopt the simplest
approach of introducing a new variable, the ``color'', by analogy with
Rothman-Keller.\cite{bib:rk}  When two particles of different color
interact we increase the conservative force, thereby increasing the
repulsion.  That is,
\begin{equation}
  \alpha \mapsto \alpha_{ij} = \cases{%
    \alpha_0 & if particles $i$ and $j$ are the same color \cr
    \alpha_1 & if particles $i$ and $j$ are different colors, \cr
    } \label{model}
\end{equation}
where $\alpha_0$ and $\alpha_1$ are constants with $0 \le \alpha_0 <
\alpha_1$.  As for the single-phase DPD fluid, the Navier-Stokes
equations are obeyed within regions of homogeneity in each of the two
immiscible fluids, while detailed balance is preserved, at least in
the limit of continuous time.\cite{coveney:multicomponent} We would
like to emphasize that with the sole exception of the choice of
finite-difference algorithm, this model is identical to that used in
our most-recently published simulations.\cite{nov-cov:binary}

\section{Evaluation criteria}\label{criteria}

A good finite-difference method should be fast, stable, accurate, easy
to implement, and require little storage.  Since for most large
systems the force evaluation dominates the computation time, the
fastest algorithm is the one that evaluates the force the fewest times
for a given accuracy.  In general, this means that stable methods
which allow large time steps are preferable.  In this context we must
be careful to distinguish the stability of a method from the stability
of the system being studied, as the dynamical systems we wish to study
are extremely unstable, with small perturbations growing
exponentially.\cite{allen:liquids,haile:md,frenkel:molecular} A stable
method will respond better to large sizes of timestep than an unstable
method.  The accuracy of an algorithm is difficult to measure in
practice because of the extreme instability of our dynamical systems.
Practical accuracy is best quantified by requiring a method to give a
trajectory in phase space that is physical for the system (i.e.\ 
similar to that observed in experiments), and for which calculated
properties of the system are close to their theoretical values.  An
idea of the accuracy of a method can also be obtained by considering
the local truncation error, expressed in terms of the order of the
timestep occurring in the first truncated term of the corresponding
Taylor series.  Because the forces considered are not smooth, low
order algorithms may possibly be more accurate than high order
algorithms.

What criteria can we use to evaluate the suitability of a method?  The
literature\cite{allen:liquids,haile:md,frenkel:molecular} claims that
a good finite-difference method will be area preserving, time
reversible, energy conserving, and ergodic.  The first three criteria
are closely related.  We need to be careful in distinguishing between
the properties of the finite-difference algorithm and those of the
system it is solving.  For example, when we describe an algorithm as
being time reversible, we are not saying that it corrects or
compensates for the lack of time symmetry in a non-conservative
system, but rather that when applied to a conservative system this
finite-difference algorithm is able to retrace the precise sequence of
steps in the backward evolution which it traversed originally in the
forward time direction.  A similar confusion may arise when we
consider ergodicity.  Is this not a property normally associated with
the rules of interaction of systems?  What does it mean to say that a
finite-difference algorithm is ergodic?  When we describe an algorithm
as being ergodic, we mean that an ergodic system updated by this
algorithm will still sample the full phase space; a non-ergodic
algorithm updating the same system will not.

We also need to consider the relevance of these four criteria to
non-conservative forces.  For example, is there any use in having an
algorithm which conserves energy when our DPD interactions conserve
only momentum?  We argue that unnecessary losses are indeed
undesirable; a finite-difference algorithm which loses energy by
consistently underestimating velocity would exhibit lower temperature
than a similar algorithm which does not.

An algorithm which preserves the volume of phase space will have a
unit Jacobian for the transformation of variables from one timestep to
the next.  Time reversibility can also be checked analytically
(ignoring for the moment the properties of the forces).  Energy
conservation normally follows from time reversibility, but it can also
be verified for a conservative system, such as the (one-dimensional)
simple harmonic oscillator
\begin{equation}
  \a = -x,
\end{equation}
where the total energy is $E = (x^2 + \v^2)/2$.

Ergodicity can be checked by calculating the equilibrium temperature
and pressure of an homogeneous fluid, and comparing the velocity
distribution with a normal (Gaussian) distribution.\cite{haile:md} As
this is not a sufficient condition, in the present paper we also
simulate a binary immiscible fluid using each of the algorithms,
testing Laplace's law and comparing the rate of growth of domains with
the extensive experimental and theoretical literature, and with our
earlier simulations\cite{nov-cov:binary,cov-nov:binary} using the
standard DPD finite-difference algorithm (see Section~\ref{euler}).
It has also been noted in DPD simulations of an ideal gas (i.e.\ 
$\alpha = 0$) using this algorithm that the radial pair-correlation
function differs from the theoretical result.\cite{marsh:properties} A
good finite-difference algorithm would remove this inconsistency, so
we verify this as well.

\section{Algorithms considered for continuous-time DPD}
\label{algorithms}

For molecular dynamics simulations, the literature recommends either
one of the Verlet schemes or a multi-value
predictor-corrector.\cite{allen:liquids,haile:md,frenkel:molecular}
Runge-Kutta and extrapolation methods are considered unsuitable
because they typically require a large number of force evaluations per
timestep, which is inefficient for a many-particle system.  For the
same reason, methods which consider the second order differential
equation directly are usually preferred to methods which break it into
two first order equations.  

We do not consider implicit methods (in which a variable appears on
both sides of a single equation) since they are computationally
infeasible for any system large enough to be of interest.  This is so
because the simplest implicit method for a system of $N$ particles
requires solution of $N$ interdependent linear equations taking
O($N^3$) operations,\cite{press:nr} compared with the O($N$)
operations of an explicit method.  The need to solve these
interdependent equations also makes implicit methods much more
complicated than explicit methods.  However, where the differential
equations are very stiff (have several greatly differing time scales)
the greatly-increased algorithmic stability could outweigh the
disadvantages.  Having had reasonable results with explicit methods,
we do not believe our system is so stiff as to require implicit
schemes.

We have obtained good results by starting our simulations with all
particles at rest, with the initial position of each particle
typically selected from a random variable uniformly distributed across
the simulation space.  Particles are assumed to have been at rest for
some time, so prior positions are the same as initial positions, and
prior velocities and higher derivatives are all zero.  In our notation
$x_n = x(t)$ and $x_{n+1} = x(t+h)$, the initial conditions at $t=0$
correspond to $n=0$ while negative values of $n$ represent values
prior to the start of the simulation.  To give an idea of the relative
accuracy of each algorithm we report the local truncation error in
position as the order of the timestep appearing in the first term
omitted from the Taylor series.

\subsection{Basic methods}\label{euler}

The original algorithm proposed by Hoogerbrugge and
Koelman\cite{hoogerbrugge:dpd} is a modified Euler scheme
\begin{eqnarray}\label{HK}
  \a_n &=& f \left( x_n,~\v_n,~h \right) \cr
  \v_{n+1} &=& \v_n + h \a_n \cr
  x_{n+1} &=& x_n + h \v_{n+1} \cr
          &=& x_n + h\left(\v_n + h \a_n\right).
\end{eqnarray}
As much research has been published using this area-preserving
scheme,
\cite{hoogerbrugge:dpd,boek:colloidal,schlijper:polymer,nov-cov:binary,cov-nov:binary,marsh:properties,koelman:hard-sphere}
it is important to compare it with the other methods.  The local
truncation error is O($h^2$).

\subsection{Verlet-based methods}

Due to the velocity-dependent nature of the force
(Eq.~(\ref{dpd_force})), the traditional O($h^3$) time-reversible
velocity-Verlet algorithm
\begin{eqnarray}\label{v-Verlet}
  x_{n+1} &=& x_n + h \left( \v_n + \frac{h}{2} \a_n \right) \cr
  \a_{n+1} &=& f\left(x_{n+1}\right) \cr
  \v_{n+1} &=& \v_n + \frac{h}{2} \left( \a_n + \a_{n+1} \right)
\end{eqnarray}
cannot be used.  Ferrario\cite{allen:chemical} has suggested two
modified schemes to get around this problem.  The first is
\begin{eqnarray}\label{FerrarioI}
  \a_n &=& f\left(x_n,~\v_n,~h\right) \cr
  x_{n+1} &=& 2 x_n - x_{n-1} + h^2 \a_n \cr
  \v_n &=& {x_{n+1} - x_{n-1} \over 2 h} \cr
  \v_{n+1} &=& \v_{n-1} + 2 h \a_n
\end{eqnarray}
and the second
\begin{eqnarray}\label{FerrarioII}
  x_{n+1} &=& x_n + h \left( \v_n + \frac{h}{2} \a_n \right) \cr
  \tv &=& \v_{n-1} + 2 h \a_n \cr
  \a_{n+1} &=& f\left(x_{n+1},~\tv,~h\right) \cr
  \v_{n+1} &=& \v_n + \frac{h}{2} \left( \a_n + \a_{n+1} \right).
\end{eqnarray}
These two O($h^3$) methods are hereafter referred to as Ferrario~I
(Eq.~(\ref{FerrarioI})) and Ferrario~II (Eq.~(\ref{FerrarioII}))
respectively.  Groot and Warren\cite{groot:bridging} recently proposed
a different velocity estimate for use with the velocity-Verlet
algorithm.  Their method is
\begin{eqnarray}\label{GW}
  x_{n+1} &=& x_n + h \left( \v_n + \frac{h}{2} \a_n \right) \cr
  \tv &=& \v_n + \lambda h \a_n \cr
  \a_{n+1} &=& f\left(x_{n+1},~\tv,~h\right) \cr
  \v_{n+1} &=& \v_n + \frac{h}{2} \left( \a_n + \a_{n+1} \right).
\end{eqnarray}
This is conceptually similar to the leap-frog form of the Verlet
algorithm, for if $\lambda \ne 1$ the force is evaluated when the
positions and velocities are out of phase.  A reasonable range for
$\lambda$ would be from $0$ to $1$; we consider only \lhf{} and
\lon{}.  For conservative forces Eq.~(\ref{GW}) reduces to
Eq.~(\ref{v-Verlet}); otherwise with \lon{} its local truncation error
is O($h^3$), and O($h^2$) for $\lambda \ne 1$.

\subsection{Runge-Kutta methods}

In order to confirm the poor performance of Runge-Kutta schemes, we
consider a two-stage explicit Runge-Kutta method of the form proposed
by Gear\cite{gear:odes} for the direct solution of second order
differential equations:
\begin{eqnarray}\label{RK}
\tf_1 &=& f\left(x_n,~\v_n,~\frac{2}{3}h\right) \cr
\tx &=& x_n + \frac{2}{3} h \left( \v_n + \frac{h}{3} \tf_1 \right) \cr
\tv &=& \v_n + \frac{2}{3} h \tf_1 \cr
\tf_2 &=& f\left(\tx,~\tv,~\frac{1}{3}h\right) \cr
x_{n+1} &=& x_n + h\left(\v_n + \frac{h}{4}\left(\tf_1+\tf_2\right)\right) \cr
\v_{n+1} &=& \v_n + \frac{h}{4}\left(\tf_1 + 3 \tf_2\right).
\end{eqnarray}
In applying this O($h^4$) method to our system, we must be careful to
choose the appropriate length of timestep for each force evaluation.

\subsection{Multi-value predictor-correctors}

Gear has proposed a notation for the general class of multi-value
predictor-corrector algorithms.\cite{gear:odes} The variables to be
stored from step to step are kept in a column vector, usually in
either the Nordsieck ($N$) representation
\begin{equation}
  {\bf y}_n(N) = {\left( x_n,~h \v_n,~\frac{h^2}{2}
      \a_n,~\frac{h^3}{3!} x^{(iii)}_n,~\frac{h^4}{4!}
      x^{(iv)}_n,~\ldots
    \right)}^T 
\end{equation}
or the force ($F$) representation
\begin{equation}
  {\bf y}_n(F) = {\left( x_n,~h \v_n,~\frac{h^2}{2}
      \a_n,~\frac{h^2}{2} \a_{n-1},~\frac{h^2}{2} \a_{n-2},~\ldots
      \right)}^T.
\end{equation}
The $N$-representation is more convenient for changing timestep and
order, while the $F$-representation simplifies calculation.
Transformation from one representation to the other is
straightforward, involving a multiplication by a transformation matrix
${\sf T}$.  Gear claims that transformation from one representation
to another does not affect the truncation error or stability of the
method, but does affect the round-off error, and amount of computation
and storage.\cite{gear:odes,footnote1}

To take a step forward in time using a predictor-corrector method, we
first multiply the column vector ${\bf y}_{n-1}$ with a matrix ${\sf
  A}$ in an explicit prediction step
\begin{equation}\label{P}
  P:\quad {\bf y}_{n,0} = {\sf A} {\bf y}_{n-1},
\end{equation}
and then the forces are evaluated at the positions and velocities in
this vector,
\begin{equation}\label{E}
  E_m:\quad \a_{n,(m+1)} = f\left(x_{n,m},~\v_{n,m},~h\right).
\end{equation}
A corrected vector is obtained by adding a multiple of the difference
between the predicted and the calculated force to the existing vector
\begin{eqnarray}\label{C}
  C_m:\quad {\bf y}_{n,(m+1)} &=& {\bf y}_{n,m} 
  + {\bf l} \frac{h^2}{2} \left(\a_{n,(m+1)} - \a_{n,m}\right).
\end{eqnarray}
It is possible to iterate the force evaluation ($E$) and correction
($C$) steps after the initial prediction ($P$), leading to a
$P{(EC)}^q$ or $P{{(EC)}^q}E$ algorithm.  The subscript $m$ denotes
each evaluation-correction step from the first ($m=0$) through the
last ($m=q-1$, i.e.\ $0 \le m < q$).  It can be shown that in the
limit of a large number of iterations ($q \gg{} 1$), these algorithms are
time-reversible.\cite{haile:md}

The predictor matrix ${\sf A}$ is usually chosen to be the Pascal
triangle matrix
\begin{equation}\label{pred_matrix}
  {\sf A}(N) = \left(
    \begin{array}{ccccc}
      1 & 1 & 1 & 1 & \cr
      0 & 1 & 2 & 3 & \cr
      0 & 0 & 1 & 3 & \cr
      0 & 0 & 0 & 1 & \cr
        &   &   &   & \ddots 
    \end{array}
  \right)
\end{equation}
(in the $N$-representation), which means that Eq.~(\ref{P}) predicts
according to the familiar Taylor series expansion.  The column vector
${\bf l}$ is chosen by accuracy and stability arguments.  The
corrector vectors for 3- and 4-value methods\cite{footnote2} to solve
second order differential equations in the $N$-representation are
\begin{equation}
  {\bf l}(N) = {\left( \frac{1}{3},~1,~1 \right)}^T
\end{equation}
and
\begin{equation}
  {\bf l}(N) = {\left( \frac{1}{6},~\frac{5}{6},~1,~\frac{1}{3}
    \right)}^T,
\end{equation}
giving local truncation error of O($h^4$) and O($h^5$) respectively.
The correction vectors are identical in the $F$-representation except
that all components beyond the third are
zero.\cite{gear:odes,berendsen:algorithms}

We would like to consider these two methods further modified by the
insertion of a factor $\lambda$ (typically ranging from $0$ to $1$)
into the velocity prediction:
\begin{equation}\label{pred_matrix_l}
  {\sf A}(N) = \left(
    \begin{array}{cccc}
      1 & 1 & 1 & 1 \cr
      0 & 1 & 2\lambda & 3 \cr
      0 & 0 & 1 & 3 \cr
      0 & 0 & 0 & 1 \cr
    \end{array}
  \right),
\end{equation}
(this matrix being for the 4-value method).  While this predicted
value of velocity will be used for the first force evaluation, the
Taylor-predicted velocity will be used in the correction step and
subsequent iterations so that the methods will benefit from the
modified velocity estimate and yet will still converge to the correct
solution.  It can be shown that both Eqs.~(\ref{pred_matrix}) and
(\ref{pred_matrix_l}) are area-preserving transformations.  Written
explicitly in the $P{(EC)}^q$ $F$-representation, we have
\begin{displaymath}
  P:\quad \left\{ \begin{array}{rcl}
        x_{n,0} &=&\displaystyle x_{n-1} + h\left(\v_{n-1} +
          \frac{h}{2}\a_{n-1}\right) \smallskip\cr
        \v_{n,0} &=& \v_{n-1} + h \a_{n-1} \smallskip\cr
        \a_{n,0} &=& \a_{n-1}
      \end{array} \right. 
\end{displaymath}
\begin{displaymath}
  E_m:\quad \a_{n,(m+1)} = \cases{%
    f\left(x_{n,0},~\v_{n-1} + \lambda h\a_{n-1},~h\right)
      & if $m = 0$ \smallskip\cr
    f\left(x_{n,m},~\v_{n,m},~h\right)
      & if $m > 0$
    }
\end{displaymath}
\begin{equation}\label{Gear3}
  C_m:\quad \left\{ 
    \begin{array}{rcl}
      x_{n,(m+1)} &=&\displaystyle x_{n,m} + \frac{h^2}{6} \left(
        \a_{n,(m+1)} - \a_{n,m} \right) \medskip\cr
      \v_{n,(m+1)} &=&\displaystyle \v_{n,m} + \frac{h}{2} \left(
        \a_{n,(m+1)} - \a_{n,m} \right)
    \end{array} \right. 
\end{equation}
and 
\begin{displaymath}
  P:\quad \left\{ \begin{array}{rcl}
        x_{n,0} &=&\displaystyle x_{n-1} + h\left(\v_{n-1} + \frac{h}{6}
          \left( 4\a_{n-1} - \a_{n-2} \right)\right) \smallskip\cr 
        \v_{n,0} &=&\displaystyle \v_{n-1} + \frac{h}{2} \left( 3\a_{n-1} -
          \a_{n-2} \right) \smallskip\cr
        \a_{n,0} &=& 2\a_{n-1} - \a_{n-2} \smallskip\cr
        \a_{(n-1),0} &=& \a_{n-1}
      \end{array} \right. 
\end{displaymath}
\begin{displaymath}
  E_m:\quad \a_{n,(m+1)} = \cases{\displaystyle
    f\left(x_{n,0},~\v_{n-1} + \lambda\frac{h}{2}\left(3\a_{n-1} -
        \a_{n-2}\right),~h\right) 
      & if $m = 0$ \smallskip\cr
    f\left(x_{n,m},~\v_{n,m},~h\right)
      & if $m > 0$
    }
\end{displaymath}
\begin{equation}\label{Gear4}
  C_m:\quad \left\{ 
    \begin{array}{rcl}
      x_{n,(m+1)} &=&\displaystyle x_{n,m} +
      \frac{h^2}{12}\left(\a_{n,(m+1)} - \a_{n,m}\right) \medskip\cr
      \v_{n,(m+1)} &=&\displaystyle \v_{n,m} +
      \frac{5}{12}h\left(\a_{n,(m+1)} - \a_{n,m}\right) \smallskip\cr
      \a_{(n-1),0} &=& \a_{(n-1),m}
    \end{array} \right. 
\end{equation}
for the 3- and 4-value methods respectively.

\section{Performance of the algorithms}

In this section we report on our evaluation of the finite difference
algorithms in the previous section, our evaluation criteria being
briefly outlined in Section~\ref{criteria}.  We first present the
results of the simple evaluations checking conservation of energy and
ergodicity, before presenting more extensive results on phase
separation and surface tension.

\subsection{Energy conservation}

A system consisting of a single simple harmonic oscillator was allowed
to evolve for a thousand timesteps from an initial configuration with
unit position ($x=1$).  Two sizes of timestep were considered: one
tenth and one hundredth the period of the oscillator.  The variation
of the energy $E = (x^2 + \v^2)/2$ relative to the initial energy
$E=\frac{1}{2}$ for each of these sets of simulations is shown in
Tables~\ref{SHO1} and~\ref{SHO2} respectively; results are identical
for \lhf{} and \lon{}.  Where a median and standard deviation are
given, the energy varies periodically; where an exponent is given, the
energy grows or decays as $a \exp(b t)$.  From these tables we can see
that Gear-3 ($q=1$) and Runge-Kutta have an undesirable continual
energy loss for both sizes of timestep.  The other Gear methods show a
periodic variation of energy for the small timestep, but display a
long-term energy drift for the large timestep.  Both Ferrario methods,
Groot-Warren, and Hoogerbrugge-Koelman only show periodic variation in
energy, with Hoogerbrugge-Koelman showing by far the largest magnitude
of variation.  However, none of the other methods include the correct
(unit) relative energy within the range of their variation, and so we
cannot conclude that one periodically varying method is significantly
better than any other.

\subsection{Ergodicity}\label{ergodicity}

Two-dimensional DPD systems of 1600 identical particles were allowed
to evolve to equilibrium (roughly $t = 50$), and then the temperature
$k_B T$ and pressure $P$ were observed for the same length of time
again.  This was repeated three times for each situation from
different random initial configurations, and the results averaged.
The exact parameters used in these simulations are shown in
Table~\ref{parameters}.  The units of these parameters are those
natural to the DPD simulation; although no one has yet related them to
physical scales of length or time, it is possible to do so if one
desires to apply the simulation technique to describe a real fluid.
These particular values were suggested by Hoogerbrugge and Koelman.

From statistical mechanics, we know that the instantaneous pressure
($P$) of a system in terms of the internal virial ($W$) and the
instantaneous temperature ($T$) are given by
\begin{equation}\label{pressure}
  P = {N k_B T + W \over V} , 
\end{equation}
where
\begin{equation}
  N k_B T = {1 \over 2} \sum_i m_i \left| {\bf \v}_i \right|^2 ,
\end{equation}
and
\begin{equation}\label{virial}
  W = {1 \over 2} \sum_i \sum_{j>i} {\bf r}_{ij} \cdot{} {\bf F}_{ij} ,
\end{equation}
where $V$ is the physical volume of space, ${\bf r}_{ij} = {\bf x}_i -
{\bf x}_j$, and $|\cdot{}|$ indicates vector magnitude.  The
thermodynamic pressure and temperature are the time averages of the
instantaneous quantities.  Because of the work of Espa\~nol, Warren, and
Coveney,\cite{espanol:dpd,coveney:multicomponent} we can be sure that
a temperature exists and is meaningful, at least in the
continuous-time limit.  The pressure was calculated immediately after
the final force evaluation in each algorithm; the temperature was
further evaluated using the final values of velocity.  At the end of
each simulation, the velocity distribution in the $x$-direction was
examined; it was statistically indistinguishable from the Maxwellian
distribution whenever the temperature converged.

With the exception of Runge-Kutta, all the methods converge to correct
temperature and pressure as we decrease the timestep.  We believe this
error is due to the unusual timestep-dependent nature of the force,
because the Runge-Kutta algorithm divides the motion into two steps of
two-thirds and then one-third the total timestep.  Of all the
algorithms we consider, only the Runge-Kutta scheme subdivides the
timestep.  The relative proportion of the random component of the
force changes when we reduce the timestep size, and the algorithm does
not properly take this into account.

As we can see in Figs.~\ref{kT1} and~\ref{kT2} (in which $k_B T$ is
plotted relative to its theoretical value), the Groot-Warren (\lhf{})
and Gear-3 (\lhf{}, $q=1$) methods give the most accurate temperature.
Figs.~\ref{P1} and~\ref{P2} show that the pressure (in DPD units) is
most accurate for a given timestep with the Gear-3 (\lon{}, $q=2$) and
Gear-4 (\lhf{} and \lon{}, $q=2$) methods, with Gear-3 (\lhf{}, $q=1$
and $q=2$), Groot-Warren (\lhf{}), and Hoogerbrugge-Koelman following
close behind.  (Groot-Warren (\lhf{}) and Hoogerbrugge-Koelman gave
nearly identical results for pressure, so their symbols are
superimposed in Figs.~\ref{P1} and~\ref{P2}---look under the curve
with the filled squares: Gear-3 (\lhf{}, $q=1$).)~~The algorithms are
unstable for timesteps larger than those shown in Figs.~\ref{kT1}
and~\ref{P1}.

Excluding the time taken to calculate the temperature and pressure,
these simulations take 140~ms per timestep for the
Hoogerbrugge-Koelman algorithm on a 133~MHz DEC Alpha.  Because of the
large number of particles involved, the speed of the algorithms is
almost directly proportional to the number of force evaluations per
timestep: one for Hoogerbrugge-Koelman, the Ferrario methods, and
Groot-Warren; two for Runge-Kutta; and $q$ for the Gear
predictor-correctors.

\subsection{Phase separation kinetics}\label{sec:sep}

The study of growth kinetics in binary immiscible fluids has received
much attention lately (see our previous
papers\cite{nov-cov:binary,cov-nov:binary} for detailed references,
but note also some more recent
work\cite{emerton:microemulsions,osborn:lb}).  A central quantity is
the characteristic domain size $R(t)$; typically, one finds that
\begin{equation}\label{scaling}
  R(t) \sim t^{\beta}.
\end{equation}
Without hydrodynamic interactions, theory and
experiment\cite{bib:bray} tell us that the scaling exponent $\beta =
\frac{1}{3}$.  If flow effects are relevant
\begin{eqnarray}\label{exponents}
  \beta & = & \cases{
    \frac{1}{2} & for $R \ll{} R_h$ \medskip\cr
    \frac{2}{3} & for $R \gg{} R_h$ \cr
    }
  \quad\mbox{(in two dimensions)} \\ \nonumber
  \vspace{10pt} \\
   \beta & = & \cases{
    \frac{1}{3} & for early-time $R \ll{} R_d$ \smallskip\cr
    1           & for late-time $R_d \ll{} R \ll{} R_h$ \smallskip\cr
    \frac{2}{3} & for $R \gg{} R_h$ \cr
    }
  \quad\mbox{(in three dimensions),} \\ \nonumber
\end{eqnarray}
where $R_h = \eta^2 /(\rho \tau)$ is the hydrodynamic length and $R_d
= \sqrt{\eta D}$ is the diffusive length, expressed in terms of the
absolute (a.k.a.\ dynamic) viscosity $\eta$ (dimensional analysis
suggests it is not the kinematic viscosity $\nu = \eta / \rho$, as
previously thought\cite{nov-cov:binary,cov-nov:binary}), density
$\rho$, surface tension $\tau$, and diffusion coefficient $D$.

Because of the compute-intensive nature of the simulations in this and
the following section, we had time enough only to use a single
finite-difference method.  Our previously performed
simulations\cite{nov-cov:binary,cov-nov:binary} used the
Hoogerbrugge-Koelman algorithm, and of all the methods Groot-Warren
(\lhf{}) and Gear-3 (\lhf{}, $q=1$) gave the most accurate temperature
and pressure.  We chose to use the Groot-Warren algorithm over Gear-3
because of its superior energy conservation.

Two dimensional systems of 40000 particles were allowed to evolve from
a symmetric quench, the initial random configuration differing for
each of several simulations.  The model parameters are identical to
those used for the smaller simulations (see Table~\ref{parameters}).
Observing the growth of those simulations which used the Groot-Warren
algorithm (\lhf{}, $h=0.1$) indicated two regimes of growth following
Eq.~(\ref{scaling}): $\beta = \frac{1}{2}$ ($0.478 \pm{}0.004$) crossing
over to $\beta = \frac{2}{3}$ ($0.65 \pm{}0.02$) at $R = 8.7 \pm{}0.4$.
Reducing the timestep to $h=0.05$, we observe nearly identical
results: $\beta = \frac{1}{2}$ ($0.476 \pm{}0.004$) crossing to $\beta =
\frac{2}{3}$ ($0.63 \pm{}0.03$) at $R = 7.4 \pm{}0.9$.  Error ranges given are
the 68\% confidence intervals for the mean, using approximately five
simulations for each configuration.  Phase separation did not occur in
symmetric quenches with a larger timestep ($h=0.5$) for which the
algorithm is unstable.  This is not surprising, as the temperature is
undefined and diverges to infinity, so that the system does not remain
quenched.  At this size of timestep, the fluid modeled approximates a
gas, which we expect to remain in a thoroughly mixed state.

Asymmetric quenches with a 60:40 majority:minority phase (color) ratio
(Groot-Warren \lhf{}, $h=0.1$) displayed growth with $\beta =
\frac{1}{2}$ ($0.476 \pm{}0.002$ crossing to $0.547 \pm{}0.003$ at $R = 7.4
\pm{}0.7$).  Asymmetric quenches with a 70:30 ratio (Groot-Warren \lhf{},
$h=0.1$) also gave $\beta = \frac{1}{2}$ ($0.460 \pm{}0.003$).  Special
simulations were also set up, in which the conservation of momentum
was violated by 10--15\% for each interaction (Groot-Warren \lhf{},
$h=0.1$) by adding a random vector to the velocity of each particle
every timestep.  These simulations demonstrated a growth exponent of
$\beta = 0.407 \pm{}0.006$, ceasing growth entirely at $R = 7.57 \pm{}0.05$.
This suggests that momentum conservation is necessary for the viscous
mechanism of phase separation.  The observed growth exponent is close
to the theoretically-predicted $\beta = \frac{1}{3}$.

A sample log-log plot of the characteristic domain size ($R$) against
time for a single symmetric quench (Groot-Warren \lhf{}, $h=0.1$) is
shown in Fig.~\ref{BinSep:loglog}.  In this figure we can see the
initial non-algebraic growth as the system settles down from its
quench, the $\beta = \frac{1}{2}$ early-time diffusive scaling regime,
and the $\beta = \frac{2}{3}$ late-time viscous scaling regime.  The
results presented in this section are qualitatively the same as those
obtained in our earlier simulations using the Hoogerbrugge-Koelman
algorithm.\cite{nov-cov:binary,cov-nov:binary}

\subsection{Domain surface tension}

Phase separation in binary fluids depends, among other things, on the
interfacial tension which exists between the two immiscible phases.  A
further important test of our algorithms is thus to check on the
existence of a surface tension by confirming the validity of Laplace's
law using a series of bubble
simulations.\cite{nov-cov:binary,cov-nov:binary}

Our procedure is to set up a circular bubble (radius $R$) of one color
phase within the other phase, and allow the bubble to reach
equilibrium.  We then calculate the pressure inside ($r<0.7 R$) and
outside ($r>1.3 R$) of the bubble, using
Eqs.~(\ref{pressure}--\ref{virial}).  Repeating these experiments for
various size bubbles, we can verify Laplace's law
\begin{equation}
  P_{\text{in}} - P_{\text{out}} = {\tau \over R}.
\end{equation}
From our results in Fig.~\ref{Pdiff_vs_R} (Groot-Warren, \lhf{},
$h=0.1$) we can see that we have the desired linear behavior, and
estimate $\tau$, the interfacial surface tension, to be $0.33 \pm{} 0.05$.
This is larger than observed in our earlier Hoogerbrugge-Koelman
simulations,\cite{nov-cov:binary,cov-nov:binary} but because of the
greatly decreased noise with the Groot-Warren (\lhf{}) method, we are
more confident in the accuracy of the current results.

\subsection{Viscosity}

The absolute (dynamic) viscosity of this system can be estimated
theoretically from the continuous-time
viscosity\cite{marsh:properties} as $\eta = 2.8 \pm{} 0.4$, where the
error is assumed to be of the same order of magnitude as that of the
temperature.  In order to verify this estimate, we performed a series
of simulations of steady shear of an homogeneous fluid, using
Lees-Edwards periodic boundary conditions.  Because we intend to use
this value of viscosity to determine the hydrodynamic length $R_h$ for
comparison with the spinodal decomposition simulations in
Section~\ref{sec:sep}, we continue to use only the Groot-Warren
algorithm (\lhf{}).  A total of 63 simulations were performed, each
from a different random initial configuration.  Systems of both 1600
and 6400 particles were studied, six simulations at each of nine
different shear rates for the former and three simulations at each of
three distinct shear rates for the latter.  As the results from the
larger simulations gave a mean viscosity nearly identical to that of
the smaller ones, we can conclude that domain size effects do not bias
the smaller, faster simulations.  The velocity profile was calculated
for each set of parameters, and was found to be statistically
indistinguishable from linear in every case.

Analyzing these simulations led to a conclusion of $\eta = 1.94 \pm{}
0.01$.  Fig.~\ref{viscosity} displays the results; the error bars are
the 68\% confidence intervals for the mean, equally weighting each of
the simulations for each set of parameters.  Others have also found
discrepancies between theory and simulation, particularly regarding
the kinematic contribution to viscosity.\cite{marsh:properties}

Using the surface tension and viscosity calculated from our
simulations, we can estimate the hydrodynamic length $R_h = \eta^2
/(\rho \tau)$ to be $2.9 \pm{} 0.5$.  During our simulations we observed
the typical domain size of the crossover in spinodal decomposition,
which we estimate as $8\pm{}1$ (see Section~\ref{sec:sep}).  This length
is in agreement with Eq.~(\ref{exponents}), since for length scales
much less than $R_h$ we typically see a growth exponent of $\beta =
\frac{1}{3}$, while for length scales much larger than $R_h$ we
typically see $\beta = \frac{2}{3}$.  Although it would be more
reassuring if $R_h$ and our observed crossover length were more
comparable in size, we must remember that the theory leading to the
proposal of these growth exponents\cite{bib:bray} makes assumptions
about the underlying growth dynamics which break down in this region.
This leaves us without an accurate theoretical estimate of the
characteristic domain size at the crossover.

\subsection{Radial pair-correlation}

It has been observed that DPD simulations of an ideal gas (i.e.\ 
$\alpha = 0$) using the Hoogerbrugge-Koelman scheme have unusual
structure in the radial pair-correlation
function.\cite{marsh:properties} Since continuous-time DPD satisfies
detailed balance,\cite{bib:espan95} we expect the radial
pair-correlation function to have a constant unit value.  This
suggests that a good finite-difference algorithm would not display
this unusual structure.

In order to test each of the finite-difference methods under
consideration, systems of 40000 identical particles were evolved
($\alpha = 0$) from initially random configurations until they reached
equilibrium (roughly $t = 50$), at which point the radial
pair-correlation function was calculated.  Simulations were performed
for each of the finite-difference algorithms, using a timestep of $h =
0.05$ for all methods except Ferrario I and II, and Gear-4 (\lon{},
$q=1$).  These results are shown in Fig.~\ref{gr_0.05}.
Fig.~\ref{gr_0.01} shows the results of the simulations using the
Ferrario methods and Gear-4 with \lon{} and $q=1$, for which we used
$h = 0.01$ since they are unstable at the larger timestep.  For
comparison, the Gear-3 (\lhf{}, $q=1$), Groot-Warren (\lhf{}), and
Hoogerbrugge-Koelman methods were also tested with the smaller
timestep ($h=0.01$) and so are also shown in Fig.~\ref{gr_0.01}.  From
these figures we can see that none of the methods are significantly
better than the Hoogerbrugge-Koelman scheme, and three are noticeably
worse: Gear-3 (\lon{}, $q=1$), Gear-4 (\lhf{}, $q=1$), and
Groot-Warren (\lon{}).  Decreasing the timestep gives a marked
improvement, so that at $h = 0.01$ all the methods capture the
expected unit radial pair-correlation function to a good
approximation.

\section{Conclusions}

The most significant difference between the simulations reported in
this paper and those performed
previously\cite{nov-cov:binary,cov-nov:binary} is the improved
precision of the measured behavior.  We believe this is largely due to
changing the discrete-time algorithm from Hoogerbrugge-Koelman to
Groot-Warren.  Coupled with its superior performance in the
less-rigorous tests, we recommend the Groot-Warren method with \lhf{}
and timestep $h \le 0.1$ as most suitable for DPD simulations.  The
Gear-3 algorithm with \lhf{} and $q=1$ is the obvious second choice,
as it gave temperature and pressure to nearly the same accuracy as the
Groot-Warren method (\lhf{}), and is stable to a timestep nearly as
large.  Our only concern in recommending this modified Gear algorithm
is its poor energy conservation.

We suspect the main limitation to improving the existing algorithms is
their inability to properly address the stochastic component of the
forces; there is some literature on Brownian dynamics
simulations\cite{gunsteren:brownian} that may be useful in finding an
even better algorithm.  Compared with the solution of ordinary
differential equations, the numerical solution of stochastic
differential equations is a recent area of mathematics, and one in
which much more research is needed.\cite{kloeden:sdes}

\section*{Acknowledgments}

We are grateful to Peter Bladon, Bruce Boghosian, Alan Bray, Mike
Cates, Pep Espa\~nol, Simon Jury, Colin Marsh, John Melrose, and Matt
Segall for helpful discussions.  We thank NATO for a grant which has
supported this work in part, as has the EPSRC (U.K.) Grand Challenge
in Colloidal Hydrodynamics for access to the Cray T3D at the Edinburgh
Parallel Computing Centre.  KEN gratefully acknowledges financial
support from NSERC (Canada) and the ORS Awards Scheme (U.K.).


\newpage


\begin{table}[tbhp]
  \begin{tabular}{lccc}
    \tableline
    Method                             & Median & Std.\ dev.\ & Exponent \\ 
    \tableline
    Ferrario I (Eq.~(\ref{FerrarioI}))    & 1.142  & 0.085 & \\
    Ferrario II (Eq.~(\ref{FerrarioII}))  & 1.054  & 0.039 & \\
    Gear-3 (Eq.~(\ref{Gear3}), $q=1$)     &        &       & $-0.024$ \\
    Gear-3 (Eq.~(\ref{Gear3}), $q=2$)     &        &       & $0.0016$ \\
    Gear-4 (Eq.~(\ref{Gear4}), $q=1$)     &        &       & $-0.0044$ \\
    Gear-4 (Eq.~(\ref{Gear4}), $q=2$)     &        &       & $0.00017$ \\
    Groot-Warren (Eq.~(\ref{GW}))         & 1.054  & 0.039 & \\
    Hoogerbrugge-Koelman (Eq.~(\ref{HK})) & 1.107  & 0.25  & \\
    Runge-Kutta (Eq.~(\ref{RK}))          &        &       & $-0.0044$ \\
    \tableline
  \end{tabular}
  \caption{Simple harmonic oscillator energy relative to initial
    energy for an homogeneous fluid, with a timestep of one tenth the
    period.  A given median and standard deviation indicate periodic
    variation; an exponent describes exponential growth or
    decay.}\label{SHO1}
\end{table}

\begin{table}[tbhp]
  \begin{tabular}{lccc}
    \tableline
    Method                              & Median  & Std.\ dev.\ & Exponent \\ 
    \tableline
    Ferrario I (Eq.~(\ref{FerrarioI}))    & 1.00050 & 0.00042  & \\
    Ferrario II (Eq.~(\ref{FerrarioII}))  & 1.00049 & 0.00035  & \\
    Gear-3 (Eq.~(\ref{Gear3}), $q=1$)     &         &      & $-2.6\times{}10^{-6}$ \\
    Gear-3 (Eq.~(\ref{Gear3}), $q=2$)     & 0.99951 & 0.00012  & \\
    Gear-4 (Eq.~(\ref{Gear4}), $q=1$)     & 1.00032 & 0.000012 & \\
    Gear-4 (Eq.~(\ref{Gear4}), $q=2$)     & 1.00033 & 0.000013 & \\
    Groot-Warren (Eq.~(\ref{GW}))         & 1.00049 & 0.00035  & \\
    Hoogerbrugge-Koelman (Eq.~(\ref{HK})) & 1.0020  & 0.022    & \\
    Runge-Kutta (Eq.~(\ref{RK}))          &         &      & $-4.4\times{}10^{-7}$ \\ 
    \tableline
  \end{tabular}
  \caption{Simple harmonic oscillator energy relative to initial
    energy for an homogeneous fluid, with a timestep of one hundredth
    the period.  A given median and standard deviation indicate
    periodic variation; an exponent describes exponential
    decay.}\label{SHO2}
\end{table}

\begin{table}
  \begin{tabular}{cc}
    \tableline
    Model                  &               \\
    Parameter              & Value         \\
    \tableline
    $\alpha_0$             & 7.063 \\
    $\alpha_1$             & 7.487 \\
    $\gamma$               & 5.650 \\
    $\sigma$               & 1.290 \\
    $m_i$                  & 1             \\
    $r_c$                  & 1.3           \\
    $\rho$                 & 4             \\
    \tableline
  \end{tabular}
  \caption{DPD model parameters (see
    Section~\protect\ref{dpd}).}\label{parameters}
\end{table}


\begin{figure}[h]
  \vspace{12pt}
  \caption{Relative equilibrium temperature ($k_B T$) vs.\ timestep
    ($h$) for an homogeneous fluid.  Data are shown for the various
    methods: 
    ---$+$---~Ferrario I,
    \ldashed{}$\times{}$\ldashed{}~Ferrario II,
    \sdashed{}$\square$\sdashed{}~Gear-3 (\lon{}, $q=1$),
    $\cdots\blacksquare\cdots$~Gear-3 (\lhf{}, $q=1$),
    \dotdash{}$\circ$\dotdash{}~Gear-3 (\lon{}, $q=2$),
    ---$\bullet$---~Gear-3 (\lhf{}, $q=2$),
    \ldashed{}$\vartriangle$\ldashed{}~Gear-4 (\lon{}, $q=1$),
    \sdashed{}$\blacktriangle$\sdashed{}~Gear-4 (\lhf{}, $q=1$),
    $\cdots\triangledown\cdots$~Gear-4 (\lon{}, $q=2$),
    \dotdash{}$\blacktriangledown$\dotdash{}~Gear-4 (\lhf{}, $q=2$),
    ---$\lozenge$---~Groot-Warren (\lon{}),
    \ldashed{}$\blacklozenge$\ldashed{}~Groot-Warren (\lhf{}),
    \sdashed{}$\ast$\sdashed{}~Hoogerbrugge-Koelman,
    $\cdots\bigstar\cdots$~Runge-Kutta.  
    The solid line drawn at unity is for reference.
    }
\end{figure}

\begin{figure}[h]
  \caption{Relative equilibrium temperature ($k_B T$) vs.\ timestep
    ($h$) for an homogeneous fluid (detail).  Data are shown for the
    various methods:
    \ldashed{}$\times{}$\ldashed{}~Ferrario II,
    \sdashed{}$\square$\sdashed{}~Gear-3 (\lon{}, $q=1$),
    $\cdots\blacksquare\cdots$~Gear-3 (\lhf{}, $q=1$),
    \ldashed{}$\vartriangle$\ldashed{}~Gear-4 (\lon{}, $q=1$),
    \sdashed{}$\blacktriangle$\sdashed{}~Gear-4 (\lhf{}, $q=1$),
    ---$\lozenge$---~Groot-Warren (\lon{}),
    \ldashed{}$\blacklozenge$\ldashed{}~Groot-Warren (\lhf{}).
    The solid line drawn at unity is for reference.
    }
\end{figure}

\begin{figure}[h]
  \caption{Equilibrium pressure ($P$) vs.\ timestep ($h$) for an
    homogeneous fluid.  Data are shown for the various methods:
    ---$+$---~Ferrario I,
    \ldashed{}$\times{}$\ldashed{}~Ferrario II,
    \sdashed{}$\square$\sdashed{}~Gear-3 (\lon{}, $q=1$),
    $\cdots\blacksquare\cdots$~Gear-3 (\lhf{}, $q=1$),
    \dotdash{}$\circ$\dotdash{}~Gear-3 (\lon{}, $q=2$),
    ---$\bullet$---~Gear-3 (\lhf{}, $q=2$),
    \ldashed{}$\vartriangle$\ldashed{}~Gear-4 (\lon{}, $q=1$),
    \sdashed{}$\blacktriangle$\sdashed{}~Gear-4 (\lhf{}, $q=1$),
    $\cdots\triangledown\cdots$~Gear-4 (\lon{}, $q=2$),
    \dotdash{}$\blacktriangledown$\dotdash{}~Gear-4 (\lhf{}, $q=2$),
    ---$\lozenge$---~Groot-Warren (\lon{}),
    \ldashed{}$\blacklozenge$\ldashed{}~Groot-Warren (\lhf{}),
    \sdashed{}$\ast$\sdashed{}~Hoogerbrugge-Koelman,
    $\cdots\bigstar\cdots$~Runge-Kutta.
    }
\end{figure}

\begin{figure}
  \caption{Equilibrium pressure ($P$) vs.\ timestep ($h$) for an
    homogeneous fluid (detail).  Data are shown for the various
    methods:
    $\cdots\blacksquare\cdots$~Gear-3 (\lhf{}, $q=1$),
    \dotdash{}$\circ$\dotdash{}~Gear-3 (\lon{}, $q=2$),
    ---$\bullet$---~Gear-3 (\lhf{}, $q=2$),
    $\cdots\triangledown\cdots$~Gear-4 (\lon{}, $q=2$),
    \dotdash{}$\blacktriangledown$\dotdash{}~Gear-4 (\lhf{}, $q=2$),
    \ldashed{}$\blacklozenge$\ldashed{}~Groot-Warren (\lhf{}),
    \sdashed{}$\ast$\sdashed{}~Hoogerbrugge-Koelman,
    $\cdots\bigstar\cdots$~Runge-Kutta.
    }
\end{figure}

\begin{figure}
  \caption{Growth of characteristic domain size for a single symmetric
    quench (Groot-Warren \lhf{}, $h=0.1$).  Solid lines are of slope
    $\frac{1}{2}$ and $\frac{2}{3}$.}
\end{figure}

\begin{figure}
  \caption{Pressure difference as a function of bubble size
    (Groot-Warren, \lhf{}, $h=0.1$).}
\end{figure}

\begin{figure}
  \caption{Absolute viscosity ($\eta$) vs.\ shear rate for an
    homogeneous fluid.  Symbols indicate simulation size:
    $+$~$N=1600$, and $\times{}$~$N=6400$.}
\end{figure}

\begin{figure}
  \caption{Radial pair correlation function $g(r)$ vs.\ distance $r$
    for a DPD ideal gas ($\alpha=0$), with timestep $h=0.05$.  Data
    are shown for the various methods:
    \sdashed{}$\square$\sdashed{}~Gear-3 (\lon{}, $q=1$),
    $\cdots\blacksquare\cdots$~Gear-3 (\lhf{}, $q=1$),
    \dotdash{}$\circ$\dotdash{}~Gear-3 (\lon{}, $q=2$),
    ---$\bullet$---~Gear-3 (\lhf{}, $q=2$),
    \sdashed{}$\blacktriangle$\sdashed{}~Gear-4 (\lhf{}, $q=1$),
    $\cdots\triangledown\cdots$~Gear-4 (\lon{}, $q=2$),
    \dotdash{}$\blacktriangledown$\dotdash{}~Gear-4 (\lhf{}, $q=2$),
    ---$\lozenge$---~Groot-Warren (\lon{}),
    \ldashed{}$\blacklozenge$\ldashed{}~Groot-Warren (\lhf{}),
    \sdashed{}$\ast$\sdashed{}~Hoogerbrugge-Koelman,
    $\cdots\bigstar\cdots$~Runge-Kutta.
    The solid line drawn at unity is for reference.
    }
\end{figure}

\begin{figure}
  \caption{Radial pair correlation function $g(r)$ vs.\ distance $r$
    for a DPD ideal gas ($\alpha=0$), with timestep $h=0.01$.  Data
    are shown for the various methods:
    ---$+$---~Ferrario I,
    \ldashed{}$\times{}$\ldashed{}~Ferrario II,
    $\cdots\blacksquare\cdots$~Gear-3 (\lhf{}, $q=1$),
    \ldashed{}$\vartriangle$\ldashed{}~Gear-4 (\lon{}, $q=1$),
    \ldashed{}$\blacklozenge$\ldashed{}~Groot-Warren (\lhf{}),
    \sdashed{}$\ast$\sdashed{}~Hoogerbrugge-Koelman.
    The solid line drawn at unity is for reference.
    }
\end{figure}
\newpage


\setcounter{figure}{0}

\begin{figure}[p]
  ~\\
  \vspace{4cm}

  \hspace{2.5cm} \epsfbox{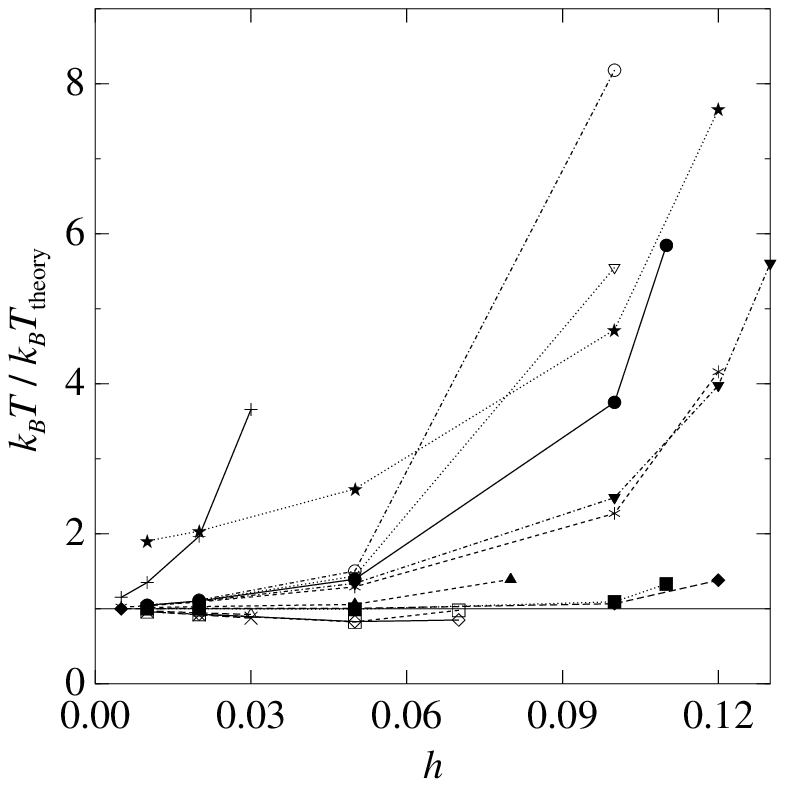}
  \vspace{7cm}
  \caption{Keir E. Novik, Journal of Chemical Physics}\label{kT1}
\end{figure}
\newpage

\begin{figure}[p]
  ~\\
  \vspace{4cm}

  \hspace{2.5cm} \epsfbox{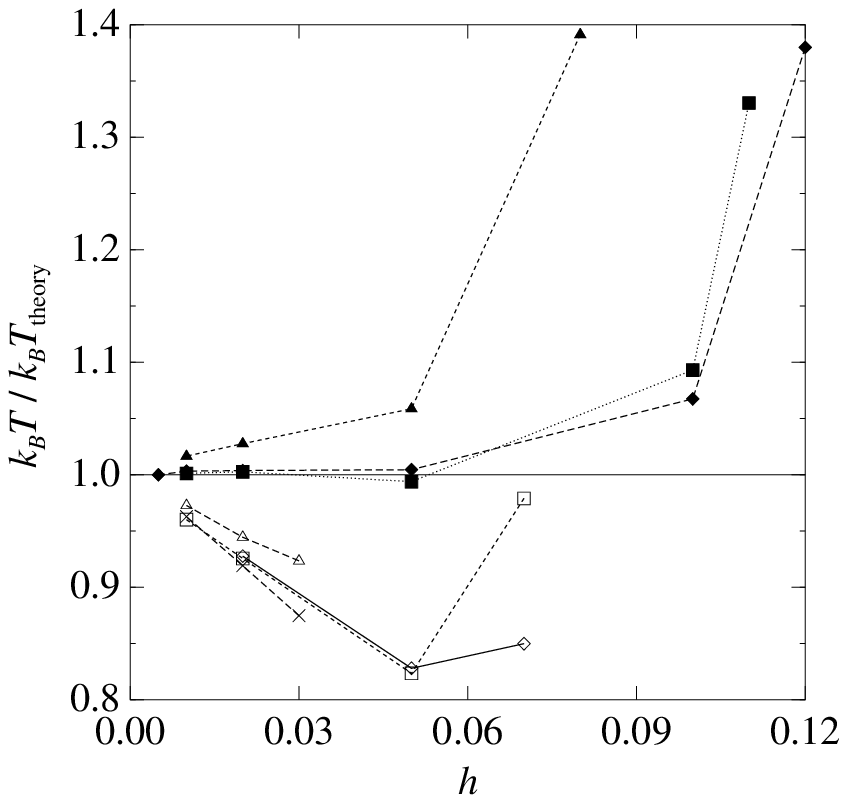}
  \vspace{7cm}
  \caption{Keir E. Novik, Journal of Chemical Physics}\label{kT2}
\end{figure}
\newpage

\begin{figure}[p]
  ~\\
  \vspace{4cm}

  \hspace{2.5cm} \epsfbox{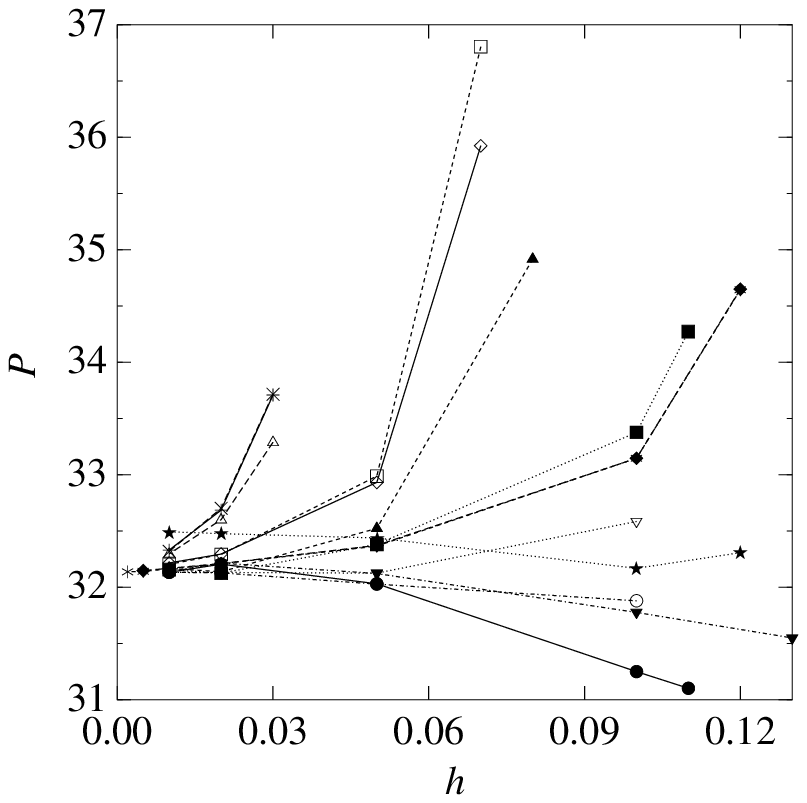}
  \vspace{7cm}
  \caption{Keir E. Novik, Journal of Chemical Physics}\label{P1}
\end{figure}
\newpage

\begin{figure}[p]
  ~\\
  \vspace{4cm}

  \hspace{2.5cm} \epsfbox{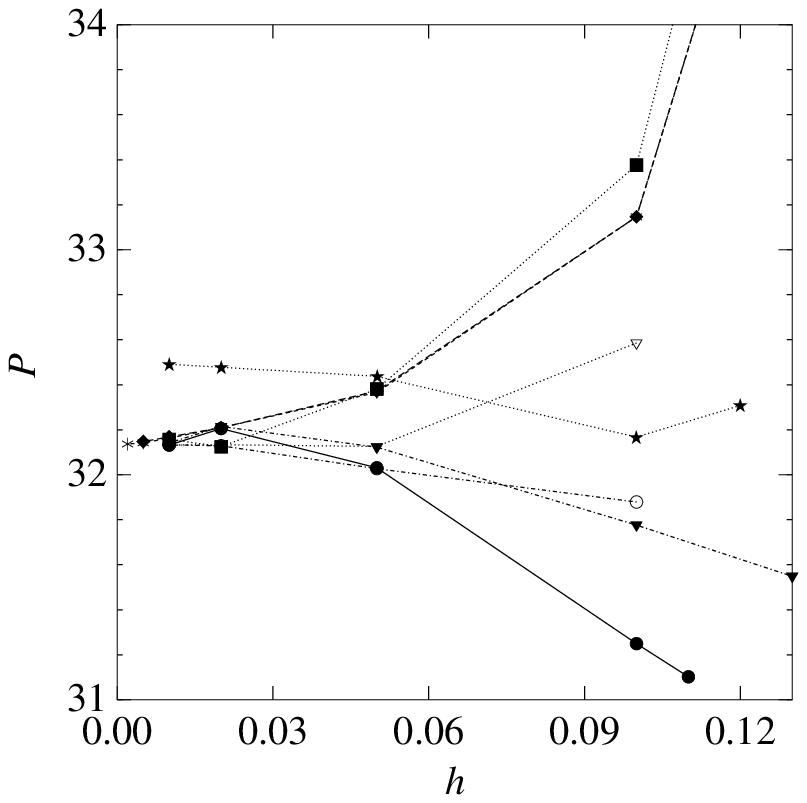}
  \vspace{7cm}
  \caption{Keir E. Novik, Journal of Chemical Physics}\label{P2}
\end{figure}
\newpage

\begin{figure}
  ~\\
  \vspace{4cm}

  \hspace{2.5cm} \epsfbox{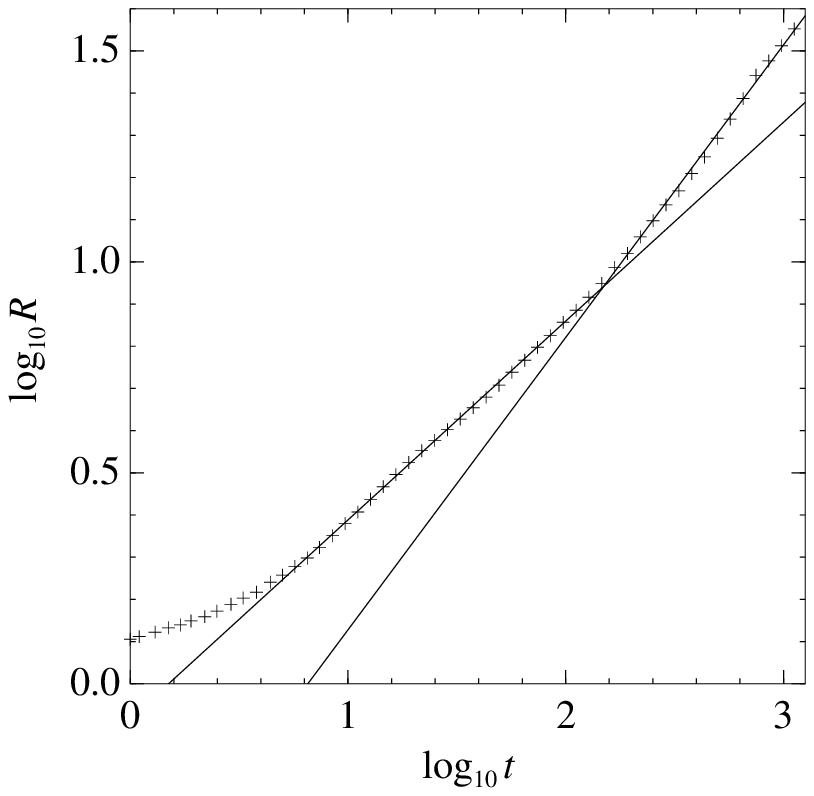}
  \vspace{7cm}
  \caption{Keir E. Novik, Journal of Chemical Physics}\label{BinSep:loglog}
\end{figure}
\newpage

\begin{figure}
  ~\\
  \vspace{4cm}

  \hspace{2.5cm} \epsfbox{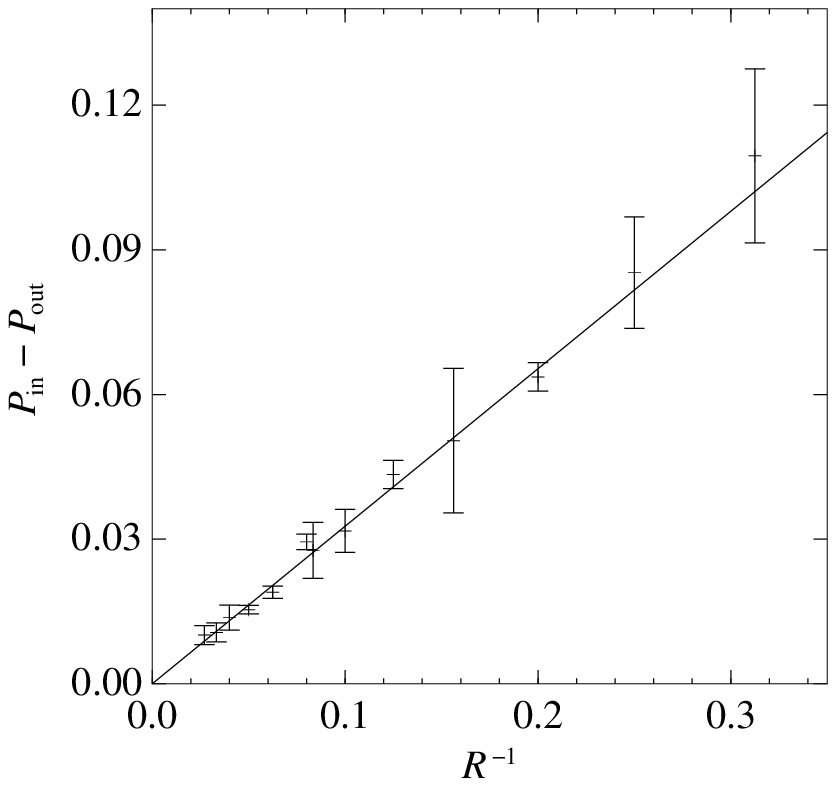}
  \vspace{7cm}
  \caption{Keir E. Novik, Journal of Chemical Physics}\label{Pdiff_vs_R}
\end{figure}
\newpage

\begin{figure}
  ~\\
  \vspace{4cm}

  \hspace{2.5cm} \epsfbox{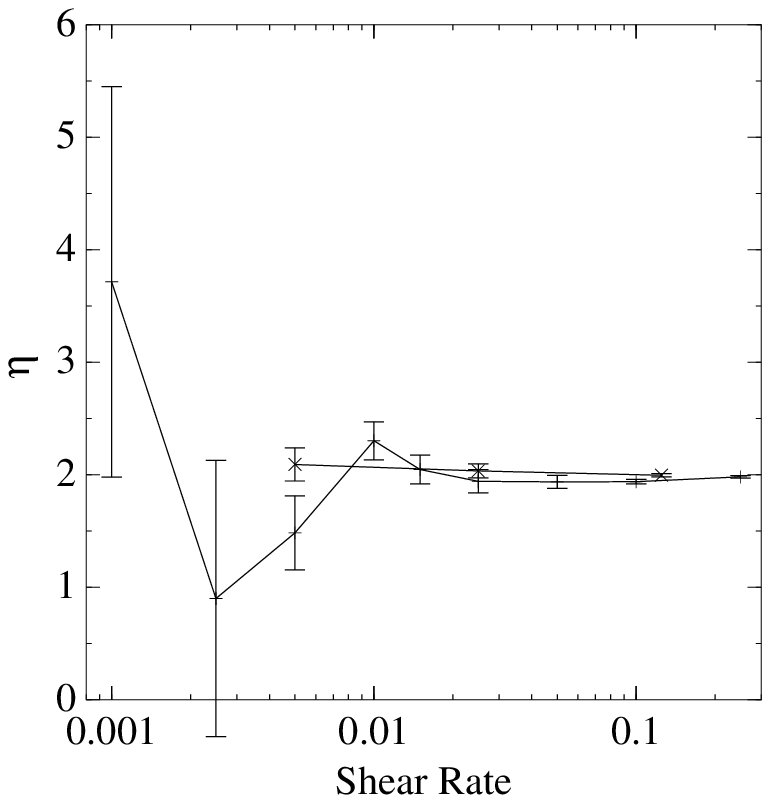}
  \vspace{7cm}
  \caption{Keir E. Novik, Journal of Chemical Physics}\label{viscosity}
\end{figure}
\newpage

\begin{figure}
  ~\\
  \vspace{4cm}

  \hspace{2.5cm} \epsfbox{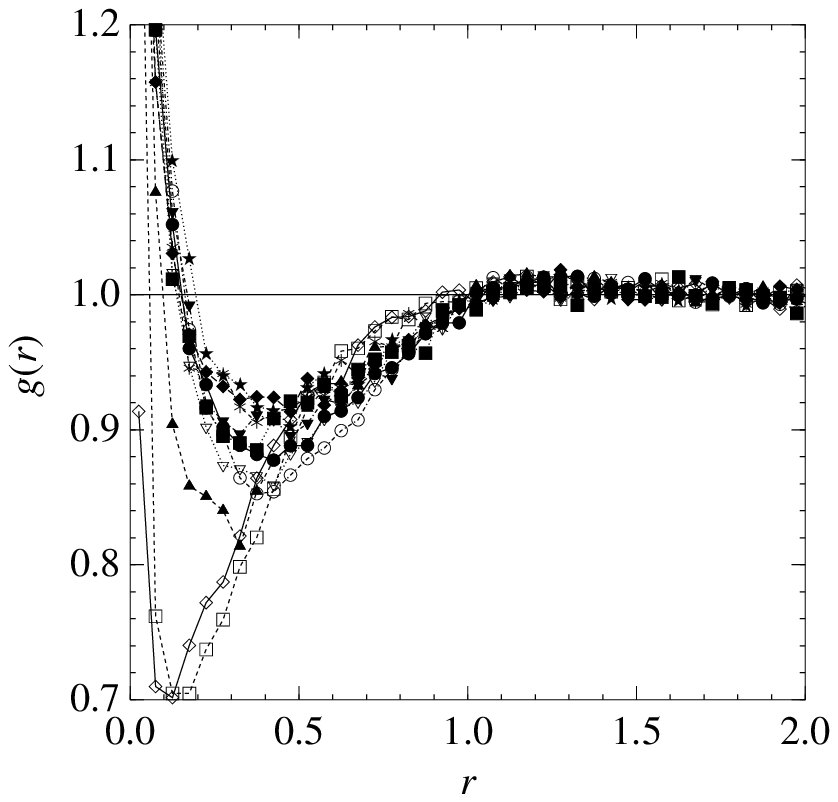}
  \vspace{7cm}
  \caption{Keir E. Novik, Journal of Chemical Physics}\label{gr_0.05}
\end{figure}
\newpage

\begin{figure}
  ~\\
  \vspace{4cm}

  \hspace{2.5cm} \epsfbox{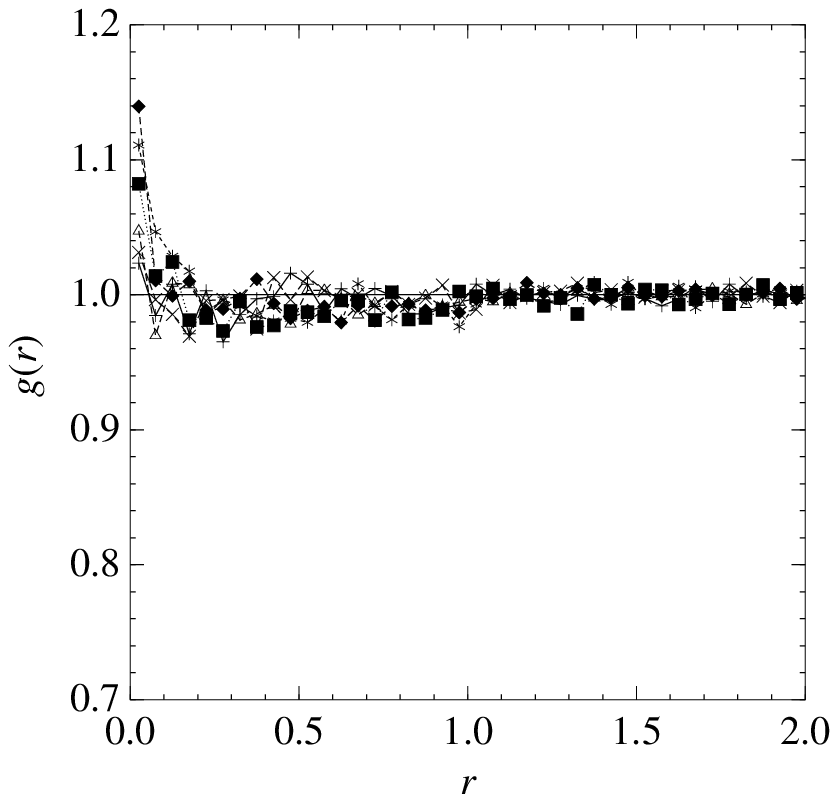}
  \vspace{7cm}
  \caption{Keir E. Novik, Journal of Chemical Physics}\label{gr_0.01}
\end{figure}


\begin{thebibliography}{99}
\bibitem{footnote1} Truncation error is the loss in accuracy from a
  finite-difference algorithm using perfect arithmetic, and is
  analogous to the error from using a truncated Taylor series to
  approximate a function of a single variable.  Round-off error is the
  precision lost by the computer due to finite precision arithmetic
  and approximations to functions.
\bibitem{footnote2} We believe this is the first time this 3-value
  corrector has been suggested, although its derivation is obvious.
\end{thebibliography}
\end{document}